\documentclass[aps,prl,twocolumn,floatfix,superscriptaddress,showpacs,amsfonts,amssymb,amsmath,preprintnumbers]{revtex4}
\usepackage{bm}
\usepackage{psfig}
\usepackage{dcolumn}

\renewcommand{\eqref}[1]{Eq.~(\ref{#1})}

\newcommand{\eqsdash}[2]{Eqs.~(\ref{#1}--\ref{#2})}

\newcommand{\exref}[1]{(\ref{#1})}

\newcommand{\Figref}[1]{Figure~\ref{#1}}
\newcommand{\figref}[1]{Fig.~\ref{#1}}
\newcommand{\figsdash}[2]{Figs.~\ref{#1}--\ref{#2}}

\newcommand{\bea}{\begin{eqnarray}}
\newcommand{\eea}{\end{eqnarray}}
\newcommand{\bal}{\begin{aligned}}
\newcommand{\eal}{\end{aligned}}
\newcommand{\bga}{\begin{gathered}}
\newcommand{\ega}{\end{gathered}}

\newcommand{\lt}{\left}
\newcommand{\rt}{\right}
\newcommand{\bl}{\bigl}
\newcommand{\br}{\bigr}
\newcommand{\la}{\langle}
\newcommand{\ra}{\rangle}

\newcommand{\const}{\text{const}}
\renewcommand{\phi}{\varphi}

\newcommand{\dd}{\partial}
\newcommand{\diff}{d}
\newcommand{\Dt}{{\diff\over\diff t}}

\newcommand{\vu}{{\bf u}}
\newcommand{\vf}{{\bf f}}
\newcommand{\vB}{{\bf B}}
\newcommand{\vb}{{\skew{-4}\hat{\bf b}}}

\newcommand{\kf}{k_0}

\newcommand{\urms}{u_\text{rms}}
\newcommand{\Brms}{B_\text{rms}}
\newcommand{\usq}{\la u^2\ra}
\newcommand{\Bsq}{\la B^2\ra}
\newcommand{\Bfr}{\la B^4\ra}

\renewcommand{\Pr}{\text{Pr}_\text{m}} 
 
\renewcommand{\Re}{\text{Re}}

\begin{document}

\preprint{{\em Phys.\ Rev.\ Lett.} {\bf 92}, 064501 (2004); 
e-print {\tt nlin.CD/0306059}}

\title{Self-Similar Turbulent Dynamo}

\author{Alexander A.\ Schekochihin}
\email{as629@damtp.cam.ac.uk}
\altaffiliation[present-time address: ]{DAMTP, 
University of Cambridge, Wilberforce Rd., Cambridge CB3 0WA, UK.}
\affiliation{Plasma Physics Group, Imperial College, 
Blackett Laboratory, Prince Consort Road, London~SW7~2BW, UK}
\author{Steven C.\ Cowley}
\affiliation{Plasma Physics Group, Imperial College, 
Blackett Laboratory, Prince Consort Road, London~SW7~2BW, UK}
\affiliation{Department of Physics and Astronomy, 
UCLA, Los Angeles, CA 90095-1547}
\author{Jason L.\ Maron}
\affiliation{Department of Physics and Astronomy, University of Rochester, 
Rochester, NY 14627} 
\affiliation{Center for Magnetic Reconnection Studies, 
Department of Physics and Astronomy, University of Iowa, 
Iowa City, IA 52242}
\author{James C.\ McWilliams}
\affiliation{Department of Atmospheric Sciences, 
UCLA, Los Angeles, CA 90095-1565}
\date{\today}

\begin{abstract}
The amplification of magnetic fields in a highly conducting fluid is
studied numerically. During growth, the magnetic field is spatially
intermittent: it does not uniformly fill the volume, 
but is concentrated in long thin folded structures. 
Contrary to a commonly held view, intermittency of the folded field 
does not increase indefinitely throughout the growth stage if 
diffusion is present. Instead, as we show, the probability-density function 
(PDF) of the field strength becomes self-similar. 
The normalized moments increase with magnetic Prandtl number 
in a powerlike fashion. 
We argue that the self-similarity is to be expected with a finite flow scale
and system size. 
In the nonlinear saturated state, intermittency is reduced and the PDF 
is exponential. Parallels are noted with self-similar behavior recently 
observed for passive-scalar mixing and for map dynamos.

\end{abstract}


\maketitle

We consider the problem of magnetic-energy amplification  
by a homogeneous isotropic turbulence in 
a conducting fluid. This effect is known as 
{\em the small-scale turbulent dynamo,} and it is important 
to the dynamics of magnetic fields 
in astrophysical objects. It is, in fact, a generic 
property of random (in time and/or space) flows that they can 
amplify magnetic fluctuations at scales smaller than the scale 
of the flow itself. The amplification is a net result of 
stretching of the field lines by the local random strain associated 
with the flow \cite{Batchelor_dynamo,Zeldovich_etal_linear,STF_book}. 
The fields generated by this mechanism have a characteristic 
structure: they concentrate in flux folds containing 
antiparallel field lines that reverse direction at the resistive 
scale and remain straight up to the flow scale 
\cite{Ott_review,SCMM_folding,SCTMM_stokes}. 
It is because of the direction reversals that the magnetic energy 
in the wave-number space is predominantly at the resistive 
scale \cite{Kazantsev,KA}. The growing folded fields also concentrate 
in small parts of the system's volume: a phenomenon called 
intermittency. 
These properties of the small-scale dynamo are most pronounced 
for systems where the fluid viscosity is much larger than 
magnetic diffusivity (magnetic Prandtl number~$\Pr=\nu/\eta\gg1$), 
i.e., where magnetic fields reverse direction at scales much 
smaller than the viscous cutoff.
This regime is realized in many astrophysical plasmas: 
examples are warm interstellar medium, intracluster and intergalactic 
plasmas. In this Letter, we study the volume-filling 
properties of the dynamo-generated fields, i.e., the distribution of the 
field strength.

Consider the equations of incompressible MHD: 
\bea
\label{NSEq}
\Dt\,\vu &=& \nu\Delta\vu - \nabla p + \vB\cdot\nabla\vB + \vf,
\quad\nabla\cdot\vu=0,\\
\label{ind_eq}
\Dt\,\vB &=& \vB\cdot\nabla\vu + \eta\Delta\vB,
\eea
where $\diff/\diff t=\dd_t+\vu\cdot\nabla$. 
The pressure~$p$ and the magnetic field~$\vB$ are normalized 
by~$\rho$ and~$(4\pi\rho)^{1/2}$, respectively, 
where $\rho=\const$ is density. 
Turbulence is excited by the forcing~$\vf$. 
The spatial scale of the forcing is usually much larger 
than the diffusion scales of the fields. 
In our simulations, we solve \eqsdash{NSEq}{ind_eq} 
by a pseudospectral method. The forcing~$\vf$ 
is chosen to be random, white ($\delta$-correlated) 
in time, and restricted to~$k/2\pi=1,2$. 
The code units are based on box size 1 and mean 
injected power $\la\vu\cdot\vf\ra=1$. 

If diffusion is ignored ($\eta=0$), 
\eqref{ind_eq} has the following formal solution 
in the comoving frame: 
\bea
\label{B_diff_free}
\ln\lt[B(t)/B_0\rt] = \int_0^t\diff t'\,\vb\vb:\nabla\vu\,(t'),
\eea
where $\vb=\vB/B$.
The Central Limit Theorem suggests that $B$ should have 
a lognormal PDF with both the mean $\la\ln B\ra\propto t$
and dispersion $D\propto t$ [cf.~\eqref{lognormal_PDF}]. 
This implies that $\la B^n\ra\propto\exp(\gamma_n t)$, where 
the growth rates depend quadratically on the order: $\gamma_n\propto n(n+3)$ 
\cite{Zeldovich_etal_linear,Chertkov_etal_dynamo,BS_metric}. 
Thus, in the diffusion-free case, the intermittency 
of the field-strength distribution increases in time 
in the sense that  the kurtosis 
$\Bfr/\Bsq^2$ and all other normalized moments 
$\la B^{mn}\ra/\la B^m\ra\la B^n\ra$ grow exponentially. 
In space, intermittency means that the growing fields do not 
uniformly fill the volume (compared to, e.g., a Gaussian 
field with the same energy). Assuming the equivalence of 
ensemble and volume averages, we may roughly interpret $\Bfr/\Bsq^2$ 
as an inverse volume-filling fraction. 

\begin{figure}[t]
\centerline{\psfig{file=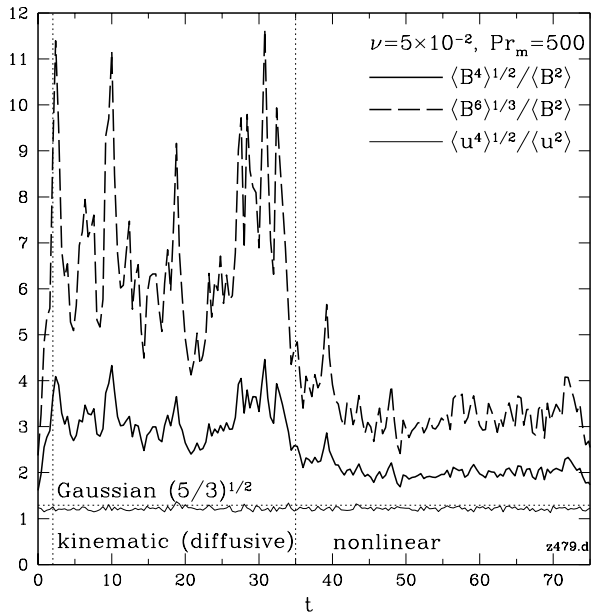,width=8.75cm}}
\caption{\label{fig_Bnt} 
Evolution of normalized moments of the magnetic-field strength. 
The (square root of) kurtosis of 
the velocity field is given for comparison. 
The kinematic (growth)
and nonlinear (saturation) stages are demarcated in the plot.} 
\end{figure}

\begin{figure*}[t]
\centerline{\psfig{file=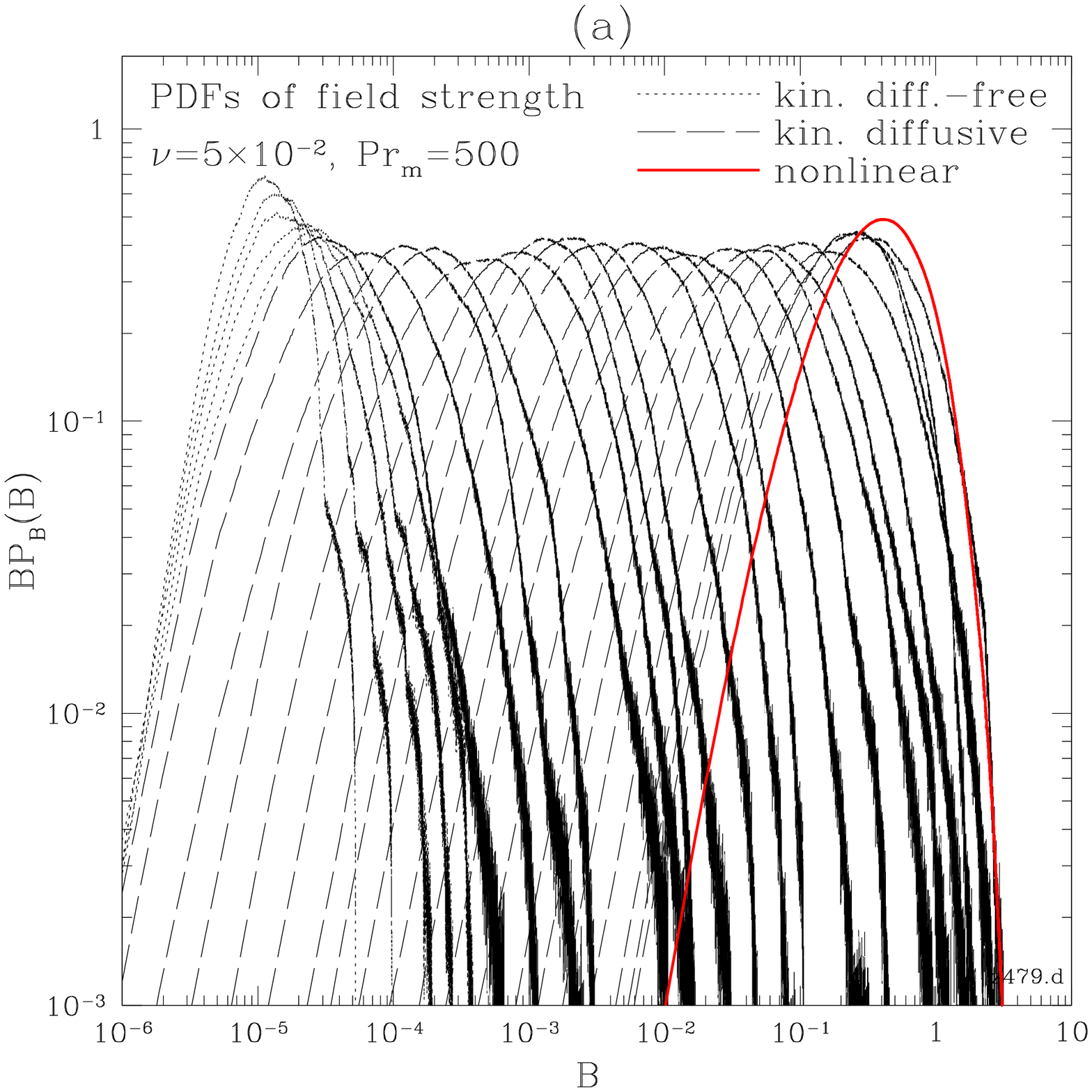,width=8.75cm}
\psfig{file=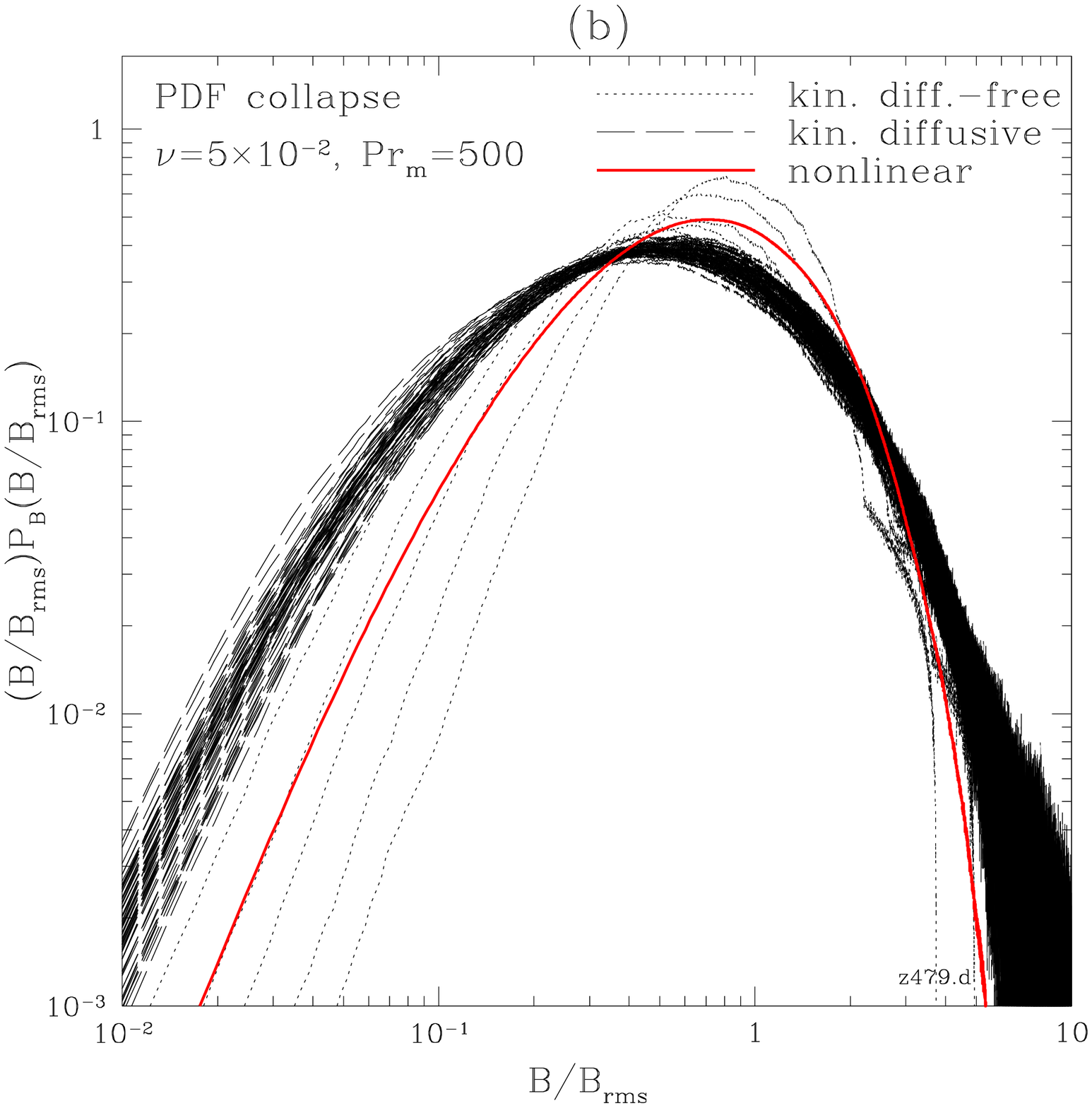,width=8.75cm}}
\caption{\label{fig_PB} 
(a) Evolution of the PDF of the magnetic-field strength. 
(b) The PDFs of~$B/\Brms$ in the kinematic regime: 
collapse onto a self-similar profile.
The PDF in the saturated state is also shown.} 
\end{figure*}

In the limit of small but finite~$\eta$, the small-scale dynamo 
can still operate, but analytical description is much harder than 
for the diffusion-free case. The traditional approach that 
makes \eqref{ind_eq} solvable in the diffusive regime 
is to consider a model random velocity field 
that is linear in space \cite{Zeldovich_etal_linear} and 
white in time \cite{Kazantsev}. The moments of $B$
can then be related to the distribution of the 
finite-time Lyapunov exponents associated with the velocity-gradient 
matrix~$\nabla\vu$, which is a function of time 
only~\cite{Zeldovich_etal_linear,Chertkov_etal_dynamo}. 
The result is that the growth rates of $\la B^n\ra$
still increase with $n$ as $\gamma_n\sim n^2$, i.e, 
intermittency continues to grow with time 
\cite{Chertkov_etal_dynamo}. 
It has so far been an accepted view that this
model adequately describes the turbulent dynamo 
in the large-$\Pr$ limit. In fact, the picture of 
increasing intermittency is {\em not} borne out by numerical 
experiments. 

In order to understand why, 
it is crucial to appreciate that the results obtained 
in the linear-velocity model apply only as long as magnetic fluctuations 
are ``unaware'' of the finiteness of the flow scale (system size).
The most intuitive, albeit nonrigorous, argument as to 
why finite-scale effects should be important is as follows. 

The fields everywhere are stretched exponentially, but with  
fluctuating stretching rates, so any occasional difference in 
field strength between different 
substructures tends to be amplified exponentially. 
Intermittency can grow with time if the system is infinite because 
for each moment $\la B^n\ra$, an ever smaller set of substructures 
can always be found in which the field has exponentially outgrown the rest 
of the system and which, therefore, dominantly contribute to $\la B^n\ra$. 
By contrast, in a finite system, 
only a finite number of exponentially growing substructures can exist, 
so the contribution to all moments must eventually come from the 
same fastest-growing one. The statistics of $B$ should, therefore, 
be {\em self-similar}, with $\la B^n\ra$ growing at rates 
proportional to~$n$, not~$n^2$, and all normalized moments 
$\la B^{mn}\ra/\la B^m\ra\la B^n\ra$ 
saturating~\footnote{Note that 
intermittency is often associated with the presence of 
sparse field structures that have disparate spatial dimensions 
(``coherent structures''). In fluid turbulence, 
these are the vortex filaments. For the dynamo-generated 
magnetic fields, the coherent structures 
are the folds, for which 
the disparate dimensions are their length 
and the field-reversal scale ($\sim$~resistive scale). 
In the linear-velocity model, the folds are elongated indefinitely. 
In reality, the length of the folds cannot be larger than 
the scale of the flow because of the bending \cite{SCMM_folding}. 
The scale separation in the folds 
is, therefore, bounded from above by~$\sim\Pr^{1/2}$.}. 

\begin{figure}
\centerline{\psfig{file=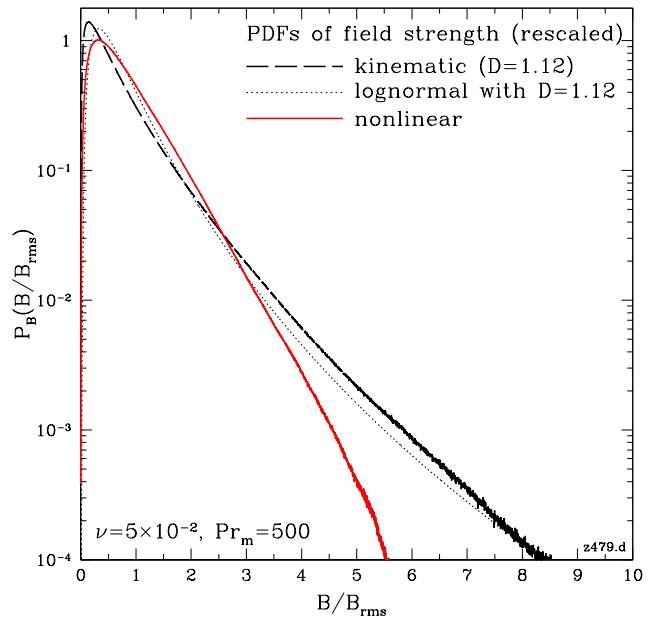,width=8.75cm}}
\caption{\label{fig_PB_tail}
The PDF of $B/\Brms$ averaged over the kinematic diffusive stage 
and the lognormal profile~\exref{lognormal_fit} with the same~$D$.
Also given is the PDF in the saturated state.}
\end{figure}

\begin{figure}
\centerline{\psfig{file=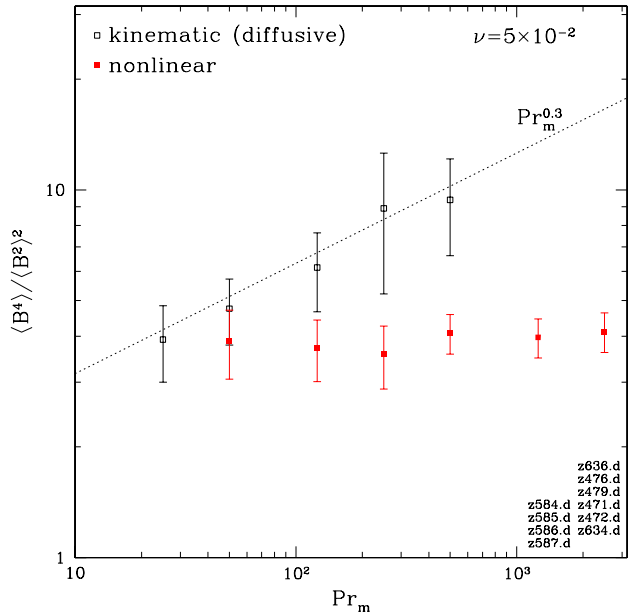,width=8.75cm}}
\caption{\label{fig_Bn_eta}
The kurtosis $\Bfr/\Bsq^2$ during the diffusive kinematic 
and the nonlinear stages of a sequence of runs with 
$\nu=5\times10^{-2}$ (low $\Re$) 
and $25\le\Pr\le2500$. All averages are over 
20 time units ($\sim$~20 turnover times).} 
\end{figure}

This is exactly what happens in our simulations. 
After initial diffusion-free growth, the normalized moments saturate 
(\figref{fig_Bnt}). The PDF of the magnetic-field 
strength becomes self-similar: namely, 
the PDFs of $B/\Brms$ (here $\Brms=\la B^2\ra^{1/2}$) collapse onto a single 
stationary profile throughout the kinematic stage of the 
dynamo (\figref{fig_PB}). 
The large-$B$ tail of the PDF of~$B/\Brms$ 
is reasonably well fitted by a lognormal distribution. 
Namely, suppose the PDF of $z=\ln B$ is 
\bea
\label{lognormal_PDF}
P_z(z)=(\pi D)^{-1/2}\exp\bl[-\bl(z-\la\ln B\ra\br)^2/D\br].
\eea 
Then $\la B^n\ra\propto\exp\lt[\la\ln B\ra n + D n^2/4\rt]$, 
so $D=\ln\bl(\Bfr^{1/2}/\Bsq\br)$. 
In the diffusive regime, $D=\const$ and 
the field-strength statistics become self-similar: 
the PDF of $\zeta=\ln(B/\Brms)$ is stationary: 
\bea
\label{lognormal_fit}
P_\zeta(\zeta)=(\pi D)^{-1/2}\exp\bl[-\bl(\zeta+D/2\br)^2/D\br].
\eea 
The lognormal fit in \figref{fig_PB_tail} is obtained by calculating 
$D$ from the numerical data and comparing the profile~\exref{lognormal_fit} 
with the numerically calculated PDF. 
The fit is qualitative but 
decent, considering the simplicity of the chosen 
profile~\exref{lognormal_fit}, large statistical errors in determining 
$\Bfr/\Bsq^2$ (\figref{fig_Bn_eta}), 
and dealiasing-induced numerical errors in resolving the field 
structure~\cite{SCTMM_stokes}. 
The PDF at low values of~$B$ appears to be powerlike 
(\figref{fig_PB}b), 
but there may be an unresolved lognormal tail at even smaller~$B$. 
Note that the large-$B$ tail describes the straight segments of the folds, 
while the small-$B$ tail 
gives the field-strength distribution for the weak fields 
in the bends \cite{SCMM_folding}. 

The plots in \figsdash{fig_Bnt}{fig_PB_tail} 
are for a $256^3$ simulation of \eqsdash{NSEq}{ind_eq} 
with $\nu=5\times10^{-2}$ and $\Pr=\nu/\eta=500$. 
The Reynolds number for this run is~$\Re=\usq^{1/2}/\nu\kf\simeq2$, 
(here $\kf=2\pi$ is the box wavenumber). 
Thus, the velocity field, while random, is smooth in space. 
This is the so-called viscosity-dominated regime, which is 
the only physical setting in which large~$\Pr$ can be resolved 
numerically. Interestingly, results for larger $\Re$ 
(``real turbulence'') and $\Pr\gtrsim1$ are very similar, 
especially in the kinematic regime. 
The reason for this is that 
the small-scale dynamo is always driven by the fastest eddies --- 
the viscous-scale ones \cite{KA}, --- 
and essentially the same field-stretching mechanism 
applies in both synthetic one-scale dynamos 
\cite{Kazantsev,Zeldovich_etal_linear,STF_book,Ott_review,SCMM_folding} 
and in turbulent systems with $\Pr\ge1$.
In runs with $\Pr=10$, $\Re\simeq100$, 
and even with $\Pr=1$, $\Re\simeq450$ (not shown), 
we have found behavior analogous to that described above: 
saturation of normalized moments in the kinematic diffusive regime 
and self-similar field-strength PDFs fairly well fitted by 
the lognormal profile~\exref{lognormal_fit}. 
In fact, the lognormal fit worked even better in these cases, 
which are somewhat less violently fluctuating, because velocity is not 
as strongly coupled to the forcing. 
More detailed comparisons will be reported in 
a future paper~\cite{SCTMM_stokes}. 

The hypothetical lognormal PDF~\exref{lognormal_PDF} 
becomes self-similar only if its dispersion~$D$ does not depend 
on time. 
In contrast, in the diffusion-free regime, $D\sim\gamma t$, 
where $\gamma\sim\nabla\vu$ is the stretching rate 
(the turnover rate of the viscous-scale 
eddies). In the case of $\eta>0$ and linear velocity field, 
the formula for $\la B^n\ra$ 
derived by Chertkov {\em et al.}~\cite{Chertkov_etal_dynamo} 
is also consistent with a 
lognormal distribution for which $D\sim\gamma t$. 
Both results are only valid transiently, during the time 
that it takes magnetic fluctuations to reach the resistive 
scale (in the former case) or the system (flow) scale (in the latter 
case). Since the scale separation is~$\sim\Pr^{1/2}$ and 
and the spreading over scales proceeds exponentially fast 
at the rate~$\sim\gamma$ (e.g.~\cite{KA}), the time during 
which intermittency increases is $t_*\sim\gamma^{-1}\ln\Pr^{1/2}$. 
Physically, this is the time necessary to form a typical fold 
with length of the order of the flow scale and 
field reversals at the resistive scale --- starting 
either from a flow-scale or a resistive-scale fluctuation. 
We might conjecture that this time determines the 
magnitude of the dispersion in the self-similar 
regime: $D\sim\gamma t_*\sim\ln\Pr^{1/2}$, which implies that 
the kurtosis~$\Bfr/\Bsq^2=\exp(2D)$ increases with~$\Pr$ in a powerlike 
fashion. The specific power law depends on prefactors that 
may be non-universal. The same holds for other normalized moments. 
\Figref{fig_Bn_eta} is an attempt to test 
this hypothesis for a sequence of simulations with increasing~$\Pr$. 
The fluctuations in the kinematic regime are very large 
(see \figref{fig_Bnt}), so 
the error bars are too wide to allow us to claim 
definite confirmation of the powerlike behavior, but our results 
are consistent with a $\Bfr/\Bsq^2\sim\Pr^{0.3}$ scaling 
\footnote{A note of caution is in order: at current resolutions, 
our lognormal fit 
is not the unique possibility consistent with numerical results: 
stretched-exponential and even steep-power-tail 
fits of comparable quality can be achieved. Note, however, that a stretched 
exponential would not be consistent with $\Pr$-dependent 
normalized moments.}. 

We emphasize that the self-similarity reported here is statistical, 
not exact. Namely, it does not imply that the magnetic field 
is simply a growing eigenmode of the induction 
equation~\exref{ind_eq}. Such an eigenmode does exist for some 
finite-scale non-random (and time-independent) flows and maps, 
owing essentially to the fact that the diffusion operator has 
a discrete spectrum in a finite domain \cite{STF_book}.
The self-similar PDF we have found here is a natural counterpart 
of this eigenmode dynamo for random flows.

Note that
for the problem of passive-scalar decay, self-similar behavior 
was also found in certain two-dimensional 
maps (the so-called ``strange 
mode''~\cite{Pierrehumbert_strange_mode,Fereday_etal,Sukhatme_Pierrehumbert,Thiffeault_Childress}), 
two-dimensional non-smooth inverse-cascading turbulence~\cite{Chaves_etal}, 
and even in scalar-mixing experiments~\cite{Rothstein_Henry_Gollub}. 
In map dynamos studied by Ott {\em et al.}~\cite{Du_Ott_grate,Ott_review}, 
moments of $B$ also grew at rates~$\propto n$ 
(these authors related such behavior to the flux-cancellation 
property of the field, i.e, to the folded structure). 
We expect that self-similar evolution is a fundamental 
property of passive advection of scalar and vector fields 
by finite-scale flows. 

A detailed discussion of the nonlinear saturation of the 
dynamo and of the magnetic-field 
intermittency in the saturated state is beyond the scope of this Letter. 
We limit ourselves to mentioning that the nonlinearity leads 
to a reduction of intermittency due to 
tighter packing of the system domain by the saturated fields 
(\figref{fig_Bnt}). 
It is not a surprising result: nonlinear back reaction imposes 
an upper bound 
on the field growth, and, once the strongest fields in the dominant 
substructure saturate, the weaker ones elsewhere have an opportunity 
to catch up. The PDF of the saturated field turns out to be 
exponential (\figref{fig_PB_tail}, 
cf.~\cite{Brandenburg_etal,Cattaneo_solar}). 
Accordingly, in the nonlinear regime, 
the kurtosis does not depend on~$\Pr$ (\figref{fig_Bn_eta}). 
Furthermore, while $\Brms$ is not $\Pr$-independent 
for finite values of $\Pr$ (though $\Brms$ does tend to a constant 
value~$\sim\urms$ as $\Pr\to\infty$), 
the PDF of $B/\Brms$ turns out to be the same 
for all $\Pr$. Further results on the nonlinear dynamo will 
be reported elsewhere~\cite{SCTMM_stokes}.

\begin{acknowledgments}
We thank S.~Boldyrev, S.~Taylor, J.-L.~Thiffeault, 
P.~Haynes, and M.~Vergassola for useful comments.
This work was supported by grants from 
PPARC (PPA/G/S/2002/00075), 
EPSRC (GR/R55344/01) 
and UKAEA (QS06992). 
Simulations were done at UKAFF (Leicester) 
and NCSA (Illinois). 

\end{acknowledgments}
\bibliography{scmm_PRL}

\end{document}